\documentclass[aps,prab,reprint,superscriptaddress,floatfix,nofootinbib,longbibliography]{revtex4-2}
\usepackage[T1]{fontenc}
\usepackage[utf8]{inputenc}
\usepackage{amsmath,amssymb,bm}
\usepackage{graphicx}
\usepackage{booktabs}
\usepackage{siunitx}
\usepackage{xcolor}
\usepackage{hyperref}
\hypersetup{colorlinks=true,linkcolor=blue,citecolor=blue,urlcolor=blue}
\begin{document}

\title{LADA: A differential algebra--based simulation package for
arbitrary-order charged-particle and light optics}

\author{Dmitry Varentsov}
\email{d.varentsov@gsi.de}
\affiliation{GSI Helmholtzzentrum f\"ur Schwerionenforschung, Darmstadt, Germany}
\author{Sergey Litvinov}
\affiliation{GSI Helmholtzzentrum f\"ur Schwerionenforschung, Darmstadt, Germany}
\author{Maksym Miski-Oglu}
\affiliation{Helmholtz Institute Mainz, 55099 Mainz, Germany}
\author{Sergiy Trotsenko}
\affiliation{GSI Helmholtzzentrum f\"ur Schwerionenforschung, Darmstadt, Germany}
\affiliation{Helmholtz-Institut Jena, Jena, Germany}

\date{26 November 2026}

\begin{abstract}
LADA computes charged-particle and light-optical transfer maps by
differential algebra in Python, with arbitrary run-time order selection.
It combines these established map methods with
ATIMA-based stopping, straggling and scattering, the GLOBAL, CHARGE and
ETACHA charge-state models, EPAX fragmentation, six-dimensional
integration through measured or computed RF fields with reference-energy
rebasing, and section-wise particle tracking with aperture and RF-window
losses. Maps retain analytic dependence on external parameters;
contraction with input moments gives beam observables and their
sensitivities without particle tracking at each setting.
Lattice import from COSY, MAD-X, Xsuite and MIRKO provides
per-record status and comparisons with available codes.
NumPy is the only required dependency. On the HHT benchmark, where the
maps of both codes agree in every coefficient through ninth order, the
optional compiled DA kernel on eight cores is 1.4 to 1.8 times faster than
COSY at orders 3 to 9. Applications include the
SIBAF irradiation facility through CALINA, proton microscopy and beam-line studies at GSI and FAIR.

\end{abstract}

\maketitle

% ---------------------------------------------------------------------------
\section{Introduction}
\label{sec:intro}

Matrix codes such as TRANSPORT~\cite{brown1980transport} compute
transfer maps at fixed low order. Differential algebra
(DA)~\cite{berz1999mmm}, implemented e.g.\ in COSY
INFINITY~\cite{makino2006cosy}, computes Taylor maps at selectable order
directly from the equations of motion.

LADA (Laser and Accelerator Differential Algebra) combines this
established method with three-dimensional RF field integration,
reference-energy rebasing, heavy-ion energy-loss, scattering and
charge-exchange models,
and particle-cloud tracking with losses. Its contribution is their
integration in a self-contained Python package with explicit
comparisons with available codes for imported lattices. Parameter-dependent
maps provide reusable response functions for nonlinear scans, matching
and sensitivity analysis. Their construction cost is paid once per
local expansion, rather than at every trial setting. NumPy is the only
required dependency.

For deterministic particle motion and geometrical light rays, the
equations of motion define a map
$\boldsymbol{u}_{\rm out}=\mathcal{M}(\boldsymbol{u}_{\rm in};\boldsymbol{\lambda})$,
where $\boldsymbol{u}$ contains phase-space coordinates and
$\boldsymbol{\lambda}$ external parameters. Its Taylor expansion about
a reference trajectory gives the transfer matrix and higher-order
aberrations. DA computes this expansion by propagating truncated
multivariate power series through the equations of motion. Particle and
light optics share this map representation~\cite{berz1999mmm,szilagyi1988}.
The applications considered here are ion optics at GSI and FAIR and
high-power laser optics (Sec.~\ref{sec:applications}); the code also
treats electrons, static magnetic fields, static or pulsed electric
fields, and periodic electromagnetic fields.

% ---------------------------------------------------------------------------
\section{The code}
\label{sec:code}

\begin{figure*}[!tp]
  \centering
  \includegraphics[width=\textwidth]{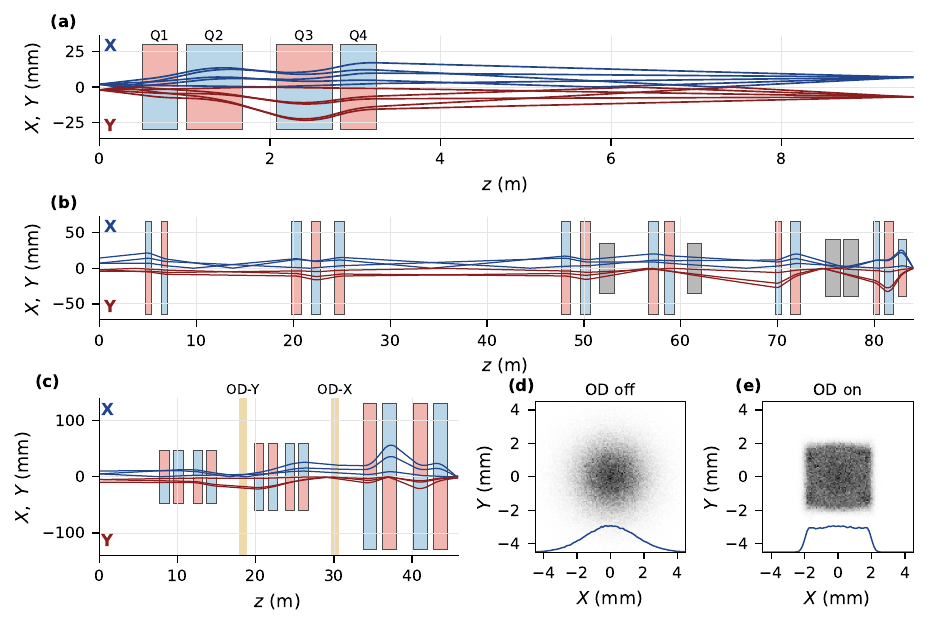}
  \caption{\label{fig:beamlines}Beam lines computed with LADA: rays from
  maps of thin slices (order 3, in (c) order 5), $|X|$ above and $-|Y|$ below the axis; magnets at
  their real half-apertures, quadrupoles
  light blue where they focus in the plane shown, light red where they
  defocus; dipoles grey.  (a)~PRIOR-II proton microscope, \SI{2.5}{GeV}
  protons at the published HHT working point~\cite{schanz2024prior}:
  point-to-point image of an object point \SI{2}{mm} off axis,
  angles up to \SI{\pm8}{mrad}.  (b)~HHT line with its final-focus
  section (FFS) at GSI, imported
  from its COSY deck, $^{238}$U$^{73+}$ at \SI{300}{MeV/u}, quadrupoles
  retuned in LADA for the energy deposition in the target at
  \SI{25}{\percent} line loss; rays on the rms ellipse of the SIS-18 beam,
  $\varepsilon_n=40\times\SI{10}{mm\,mrad}$ (4$\sigma$).  (c)~FAIR HEDgeHOB line with its
  superconducting final-focus quadruplet, U$^{28+}$ at \SI{2}{GeV/u},
  transcribed from its COSY deck, with two combined octupole/duodecapole
  elements OD-X and OD-Y (amber) and the matching section retuned in
  LADA; initial rays sample phase-space ellipses with coordinate and
  angular extents of $\pm2\sigma$.  (d,e)~Focal spot without and with
  OD-X/OD-Y: a Gaussian input is reshaped by fitting the field strengths
  to the moments of a uniform square spanning $-2$ to \SI{+2}{mm} in
  both coordinates~\cite{varentsov2005shaping}; line: horizontal profile.}
\end{figure*}

\paragraph{DA core.}
A truncated power series in $v$ variables through total degree $n$ is
stored as $\binom{n+v}{v}$ coefficients, ordered first by total degree,
then by variable powers: for two variables,
$1,x,y,x^2,xy,y^2,\ldots$. Polynomial multiplication uses a precomputed
table of output indices for monomial pairs
(e.g.\ $x^2$ and $y$ contribute to $x^2y$); products of degree above $n$
are discarded. An optional compiled DA kernel replaces NumPy multiplication
and processes independent polynomial products across cores.
Composition, inversion, elementary functions, integration and
differentiation follow Ref.~\cite{berz1999mmm}. The map order is a
run-time parameter: orders 3--5 cover routine calculations, order 7
advanced nonlinear calculations, and order 9 is the recommended upper
limit. Optimisation and validation target this range.

\paragraph{Coordinates and knobs.}
The phase-space variables are $(x,a,y,b,t,\delta)$ with $a=p_x/p_0$,
$b=p_y/p_0$, $t$ the arrival-time deviation and $\delta=\Delta E/E_0$.
Time-dependent fields use each particle's local passage time, including
its offset from the reference particle. An RF-phase mode replaces $t$
by $\omega t$.
Supported external parameters include element strengths, cavity
amplitudes and phases. Their deviations can be additional DA variables
(knobs); settings can be analytic functions of knobs expressed in DA
operations. The map is a Taylor polynomial in phase-space and knob
deviations.
Each knob acts on the element's field model, including fringes.
For $k=k_0+\Delta k$ with $k_0=0$, the magnet is off at the reference
setting, but derivatives in $\Delta k$ retain its response to switching on.
Optimisation evaluates the polynomial without re-tracking.

\paragraph{Analytic response functions.}
For an output coordinate $P(\mathbf u,\boldsymbol\eta)$, integration
over the input distribution leaves the knobs $\boldsymbol\eta$
symbolic. In particular,
\begin{equation}
 \begin{aligned}
 \mu(\boldsymbol\eta)&=\langle P\rangle,\\
 V(\boldsymbol\eta)&=\langle P^2\rangle-\mu^2(\boldsymbol\eta).
 \end{aligned}
 \label{eq:knob_moments}
\end{equation}
Contracting the coordinate powers with input moments produces explicit
polynomials in the knobs. A map of degree $n$ requires input moments
through degree $2n$ for a variance; the product $P^2$ is not truncated
back to degree $n$. Covariances follow by the same construction.
Means, rms sizes, energy spreads and emittances can thus be evaluated
from moments without transporting a particle cloud at each setting.
Moments of a known input distribution, e.g.\ a Gaussian, are analytic,
and no particles are needed at all; moments computed from a measured or
simulated particle ensemble carry its statistical error into the results.

Polynomial differentiation gives gradients, Hessians and mixed
responses to various knobs or phase-space variables. A single map therefore supplies curves
and surfaces of beam observables, local sensitivities and objectives
for numerical optimisation. These operations reuse the dependence
computed from the equations of motion, rather than fitting a response
surface to a tracking scan. Their validity is local and limited by the
map order; aperture losses and collective fields require separate
treatment. Section~\ref{sec:verification} compares a two-knob response
surface with direct RF tracking.

\paragraph{Taylor models and RDA.}
Rigorous differential algebra represents a function by a Taylor
polynomial and an interval remainder~\cite{makino2003tm}:
\begin{equation}
 \mathcal M(\mathbf u,\boldsymbol\eta)
 \in P_n(\mathbf u,\boldsymbol\eta)+I,
 \quad (\mathbf u,\boldsymbol\eta)\in D.
\end{equation}
Since the function is guaranteed to lie in $P_n+I$ everywhere in $D$,
its values can be bounded over entire coordinate and knob ranges, with applications to tolerance
analysis and exclusion of infeasible settings. LADA carries remainder
intervals through Taylor-model operations, rounds their endpoints
outward and includes polynomial coefficient rounding in the remainder;
the arithmetic is checked against exact rational calculations.
A bound for the physical transfer map additionally requires validated
field approximation and integration errors. Assigning zero remainder
to an ordinary DA map does not supply those bounds.

\paragraph{Use.}
The call \texttt{cell.map(env)} constructs the transfer map of the ordered
element list, at the order set in \texttt{env}. Here the seventh variable is
$\eta=\Delta k/k_0$, with $k=k_0(1+\eta)$ and $k_0=1.2$.
The final line extracts the numerical coefficient of $x\eta$ in
$x_{\rm out}$:
$\partial R_{11}/\partial\eta$ at the reference setting.

{\footnotesize
\begin{verbatim}
from lada import DAEnv, Lattice, Particle
from lada.elements import Drift, Quad, Sextupole

proton = Particle.proton(2500.0)  # T = 2.5 GeV
env = DAEnv(nvar=7, order=5)      # 6D + one knob
eta = env.var(6)                  # knob: dk/k of QF
cell = Lattice([
    Quad(0.2, 1.2 * (1 + eta)),  Drift(1.0),   # QF
    Sextupole(kl=0.5),           Drift(1.0),
    Quad(0.2, -1.2),             Drift(1.0)],  # QD
    particle=proton)
M = cell.map(env)                 # order 5, 7 variables
M[0].coeff(1, 0, 0, 0, 0, 0, 1)   # (x|x eta) = -0.918
\end{verbatim}}
\noindent A source deck is migrated by one command,
\texttt{lada migrate deck.fox --format cosy --pipeline trust}
(Sec.~\ref{sec:migration}).

\paragraph{RF field maps.}
Table~\ref{tab:content} lists the element classes.  Fringe fields are
integrated in DA.  Without a fringe profile a magnet, or an electrostatic
multipole, ends with the limit of a short
edge that satisfies Maxwell's equations~\cite{forest1988}; COSY's hard edge \texttt{FR 0},
which has no edge field, is an option for comparisons. The edge adds terms of
second order with the vertical coordinates for a dipole, and of order $n+1$
for a multipole of order $n$, the third for a quadrupole; given the gap, a dipole edge
also has the third-order vertical kick, which grows as the inverse gap.  The fifth-order terms of a quadrupole
edge depend on its extent and need a fringe profile.
The standard \texttt{RF} element documented in COSY INFINITY represents
a cavity as an infinitely thin, position-dependent energy kick~\cite{cosybeam102}.
LADA constructs six-dimensional Taylor maps through spatially extended
RF fields. The axisymmetric model reconstructs off-axis fields from
the on-axis electric field using
Maxwell's equations, with a radial expansion through $r^4$ in the
longitudinal field for maps of third order and higher ($r^2$ below). The 3D model instead accepts Cartesian grids of all
three electric and, optionally, magnetic field components. In its
standing-wave convention,
\begin{equation}
 \begin{aligned}
 \mathbf E(\mathbf r,T)&=\mathbf E_s(\mathbf r)\sin\Phi,\\
 \mathbf B(\mathbf r,T)&=\mathbf B_c(\mathbf r)\cos\Phi,
 \qquad \Phi=\omega T+\phi_0,
 \end{aligned}
\end{equation}
Here $T$ is the particle's physical passage time. The input quadratures
must satisfy Maxwell's equations; axial symmetry is not imposed.

For each component $F$ of $\mathbf E_s$ and $\mathbf B_c$, a transverse
polynomial is fitted on successive grid planes,
\begin{equation}
 F(x,y,z)=\sum_{i+j\leq d} f_{ij}(z)x^iy^j,
\end{equation}
and its coefficients are interpolated along $z$. Fourth-order
Runge--Kutta integration of the Lorentz equations in DA propagates all
map coefficients together, including the time and energy dependence of $\Phi$. Transit-time effects, RF focusing, phase slip
and non-axisymmetric transverse--longitudinal coupling follow from
this integration. Three approximations enter, each controlled by its own
parameter and checked by varying it alone: the off-axis field by the fit
degree $d$, the integration by the step length and the Taylor
expansion by the map order.

Coordinates retain a fixed reference within each cavity. Between
cavities, an affine change of coordinates subtracts the reference
arrival-time and energy increments and rescales momenta to the new
reference momentum, allowing successive accelerating maps to be composed.
Their linear part is the transfer matrix; higher coefficients describe
nonlinear and chromatic response.
On a generic linac module, for example, the amplitude and phase of a
final buncher are found with 132 evaluations of the knob polynomial;
the rms energy spread at the target falls from \SI{1.49}{keV/u} to
\SI{0.55}{keV/u}. For 3000 particles at third order, map construction
takes \SI{0.87}{s} and the search \mbox{0.67--\SI{0.85}{s}} on an
eight-core Apple M3 laptop with the compiled DA kernel. The search
evaluates the nonlinear maps without rebuilding them or integrating
individual particles through the RF fields.

\paragraph{Matter and cloud tracking.}
LADA implements the stopping formulas underlying ATIMA~\cite{atima}:
the Bethe stopping number
with the Lindhard--S{\o}rensen correction~\cite{lindhard1996}, computed
from the exact Dirac phase shifts, with shell and Barkas corrections and
an effective charge, and the electronic stopping of
Ziegler, Biersack and Littmark~\cite{zbl1985} below \SI{10}{MeV/u}.
Energy straggling, with the Lindhard--S{\o}rensen factor, and the
scattering angle are integrated along the energy loss, and the map of an
absorber carries the energy derivative of the whole model.  Unlike
ATIMA, the Lindhard--S{\o}rensen terms are those of a point nucleus,
and the low-energy effective charge is the variant of
Ref.~\cite{zbl1985} with the Ziegler--Manoyan screening length.  The
charge-state models GLOBAL~\cite{scheidenberger1998global},
CHARGE~\cite{scheidenberger1998global} and ETACHA97~\cite{rozet1996etacha}
are implemented in LADA's own code from the cited publications and
verified against outputs obtained by running the respective programs;
those programs are not required for LADA calculations. Fragment yields
use EPAX~\cite{summerer2012epax}.

Particle-cloud tracking propagates all six phase-space coordinates
through precomputed section maps, with aperture and RF-window losses.
For a Taylor map of any order, a particle batch is propagated by
\begin{equation}
 U_{\rm out}=V(U_{\rm in})C^{\mathsf T},
\end{equation}
where each row of $U$ contains a particle's six coordinates, $V$ contains
their monomials through the map order, and $C$ contains the map
coefficients. NumPy evaluates the monomials on whole particle arrays
and performs the matrix product using optimised BLAS where available.
Field integration is performed when constructing the maps; their reuse
avoids Runge--Kutta integration for each particle and enables high-throughput
tracking. The parallel 6D linac tracker distributes independent particle
batches across CPU processes, following the batching strategy used in
PRIORA~\cite{varentsov2026priora}.

\begin{table}[t]
\caption{\label{tab:content}Physics content of LADA.}
\begin{ruledtabular}
\begin{tabular}{p{0.24\columnwidth}p{0.70\columnwidth}}
Magnetic & drift, quadrupole, sector and rectangular dipole with
edges, multipoles to dodecapole (normal and skew), solenoid; Enge
fringe fields; 1D, 2D and 3D field maps; supplied
$B_r(r,z),B_z(r,z)$\\
Electrostatic & deflector, quadrupole, multipole, static or
time-dependent; round lens (einzel) from the axial potential\\
RF & thin cavity; on-axis field map with Maxwell expansion; 3D electric
and magnetic field grids; linac segment with reference rebasing\\
Matter & degrader, foil, window, wedge, stripper foil, gas target:
ATIMA-based stopping and straggling (Bethe with Lindhard--S{\o}rensen,
shell and Barkas corrections, density effect; ZBL below
\SI{10}{MeV/u}), scattering with Rutherford tail, charge states
(GLOBAL, CHARGE, ETACHA),
reaction cross sections, EPAX fragmentation\\
Light optics & lenses, refracting surfaces, mirrors, phase plates,
wavefront correctors; focal-spot moments, PSF\\
Analysis & Twiss, normal form, moments, optimisation, cloud
tracking with losses, isochronicity, validated Taylor models\\
\end{tabular}
\end{ruledtabular}
\end{table}

\paragraph{Performance.}
Table~\ref{tab:timing} compares 6D map computation for the HHT
final-focus line: COSY as a process and LADA within Python, on maps that
agree in every coefficient COSY prints through ninth order
(Table~\ref{tab:verification}). With a small optional compiled DA kernel
written in C, LADA is 1.4 to 1.8 times faster than COSY at orders 3 to 9
on eight cores and 1.1 times faster on one core at order 9. Products omit zero
coefficients and terms above the requested order; composition reuses
monomial products and batches independent work across cores. Elements
whose field does not depend on $s$, such as the hard-edge quadrupole
body, are mapped by the flow of their equations of motion, a Lie series
summed in DA, instead of a step-by-step integration. Results are
thread-count independent.
Composing the precomputed ESR one-turn map with itself,
$\mathcal{M}\circ\mathcal{M}$, gives the map for two revolutions. For the
ESR setting with sextupoles, this order-9 composition takes \SI{6.8}{ms} on eight
cores and \SI{16.7}{ms} on one. Product tables occupy \SI{100}{MB} at
order 9 in six variables; a slower fallback handles insufficient memory
or the absence of the compiled DA kernel.

\begin{table}[t]
\caption{\label{tab:timing}Wall time in ms for the 6D map of the HHT
line, 43 magnets and drifts (Apple M3, eight cores).  The COSY time
includes \SI{11}{ms} of process start-up; medians of 11 runs.  Both codes use COSY's hard edges.  LADA with the compiled DA kernel
on eight cores and on one core, and on the pure NumPy path.}
\begin{ruledtabular}
\begin{tabular}{rrrrr}
Order & COSY & LADA, 8 cores & 1 core & NumPy\\
 & (ms) & (ms) & (ms) & (ms)\\
\hline
3 & 15 & 11 & 11 & 40\\
5 & 24 & 17 & 17 & 172\\
7 & 74 & 46 & 56 & 1089\\
9 & 314 & 174 & 282 & 9461\\
\end{tabular}
\end{ruledtabular}
\end{table}

% ---------------------------------------------------------------------------

\section{Verification and benchmarking}
\label{sec:verification}

\begin{figure*}[!tp]
  \centering
  \includegraphics[width=\textwidth]{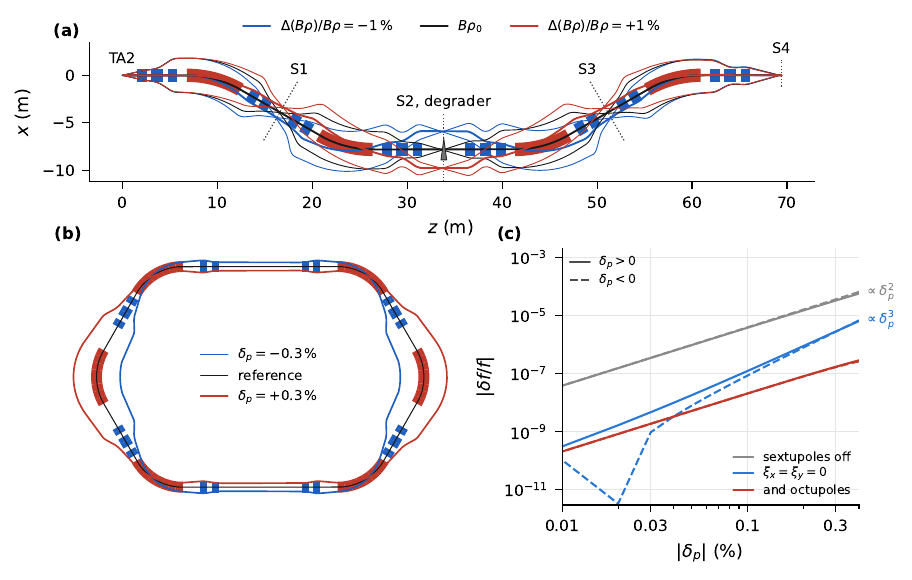}
  \caption{\label{fig:rings}(a)~Fragment separator FRS at GSI, target TA2
  to focal plane S4, from the GICOSY deck of its
  standard achromatic setting: floor plan to scale, rays of three magnetic rigidities with
  $\pm\SI{8}{mrad}$ at the target, offsets magnified $\times30$; the
  rigidities separate at the dispersive focal plane S2 and recombine at
  the achromatic focus S4.  (b)~ESR (\SI{108.36}{m}), isochronous
  setting of 2021 imported from a MIRKO deck, $\gamma=\gamma_t=1.563$
  ($^{238}$U$^{91+}$ at \SI{524}{MeV/u}): closed orbits for
  $\delta_p=\Delta p/p=\pm0.3\,\%$, offsets magnified $\times40$.
  (c)~Relative deviation of the revolution frequency on the closed orbit,
  $|\delta f/f|=|\Delta T/T_0|=|c_2\delta_p^2+c_3\delta_p^3+c_4\delta_p^4+\dots|$,
  solid $\delta_p>0$, dashed $\delta_p<0$; a term $c_n\delta_p^n$ is a
  line of slope $n$.  Sextupoles off (natural chromaticity): $c_2=3.9$; set for
  chromaticities $\xi_x=\xi_y=0$: residual
  $c_2=-0.021$ and $c_3=-105$, which cancel at
  $\delta_p=-0.02\,\%$; with thin octupoles added for $c_3=0$ (hypothetical,
  not installed): $c_2=-0.021$.}
\end{figure*}

\begin{figure}[!tbp]
  \centering
  \includegraphics[width=0.90\columnwidth]{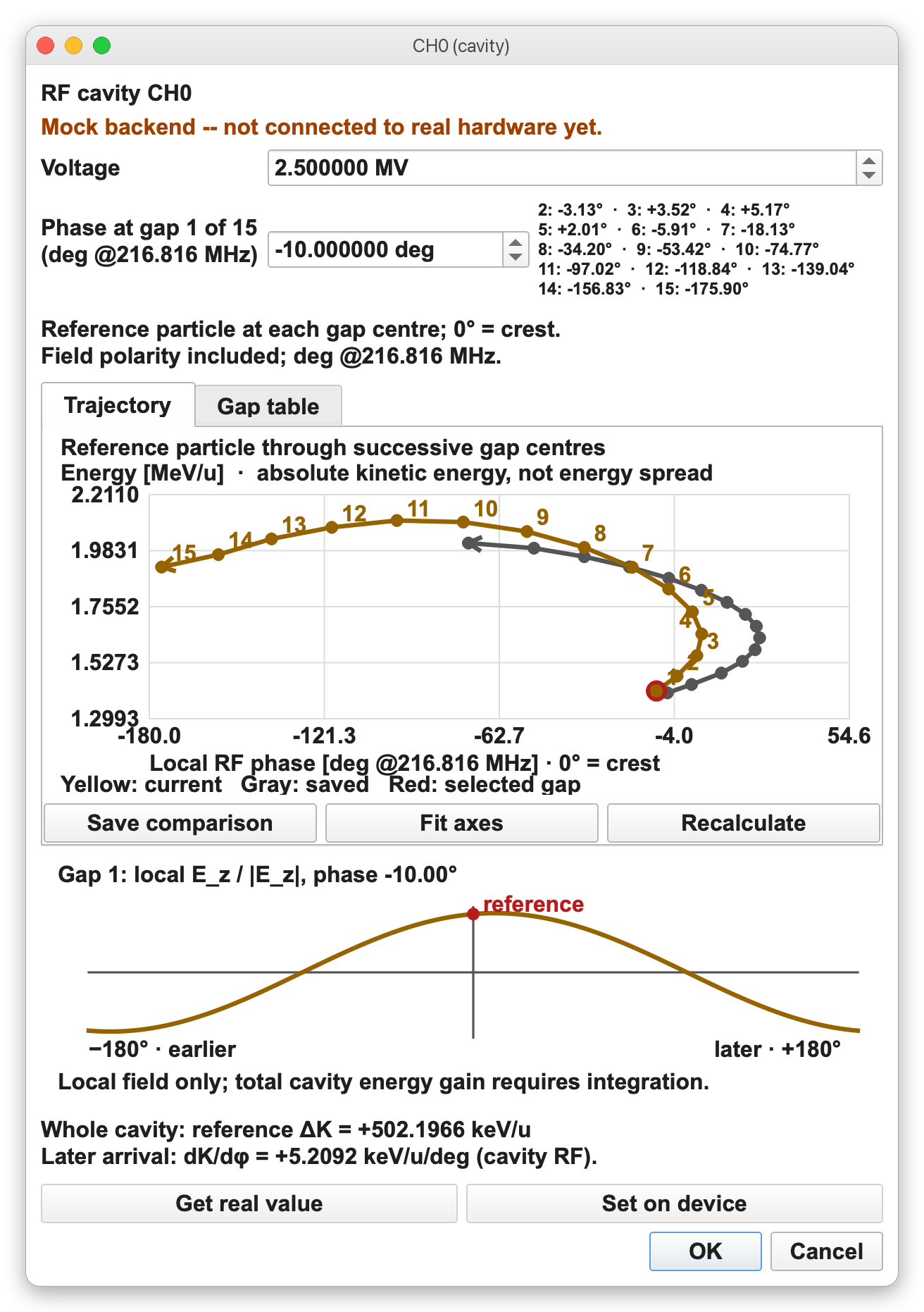}
  \caption{\label{fig:linac}Cavity dialog of Torpedo, the GUI of CALINA
  (rendered with a light print palette): SIBAF cavity CH0, He$^{2+}$, at \SI{2.5}{MV} and $-10^\circ$ at the first
  gap (dark ochre) against the startup setting (gray); red circle: gap 1.
  Kinetic energy and RF phase of the reference ion at the centres of the
  15 gaps, local field at gap 1 and the energy gain $\Delta K$ of the
  cavity, from integration through its field map with LADA.}
\end{figure}

Each element class is checked against a reference that is independent
of it (Table~\ref{tab:verification}).  Where COSY defines the result,
COSY is the reference. Tests include dipole, quadrupole, sextupole and octupole fringe fields, the full 6D map of bends with fringe
fields and curved pole faces, the time of flight through dipoles
and quadrupoles, electrostatic deflectors to second chromatic order,
einzel lenses for ions and electrons, imported lattices, and the normal
form in its stable, unstable and
parameter-dependent branches. For bends with field discontinuities,
LADA applies boundary-crossing corrections and agrees with independent
ray integration to $2\times10^{-8}$.
Agreement with a hard edge shows that two codes share a model, not that the
model is complete.  The Maxwellian edges are therefore compared with
Lorentz-force integration through a narrow edge field, with the limit of
COSY's fringe fields for a vanishing aperture, and with
PTC~\cite{forest2002ptc} on the ESR.
For the extended
RF field maps and ion--matter interaction models not covered by COSY
comparisons, the references are analytic limits, float Lorentz
integrators written independently of the DA code, other codes on
identical input (RF-Track~\cite{latina2016rftrack}), and published
data.  Energy loss, energy straggling and scattering angle of U, Xe and
C ions in FRS targets and degraders, vacuum windows, strippers and ESR
gas targets agree with CATIMA~\cite{catima}, a reimplementation of
ATIMA, within \SI{0.6}{\percent} from 5 to \SI{1000}{MeV/u} and
\SI{3}{\percent} at \SI{4}{MeV/u}. In contrast, the relativistic Bohr
formula evaluated at the ion's energy before entering the absorber
underestimates the rms energy spread by about \SI{40}{\percent} for
U$^{92+}$ at \SI{1}{GeV/u} in \SI{2}{g/cm^2} of Al:
\SI{0.245}{MeV/u} versus \SI{0.412}{MeV/u} from CATIMA.  The exception is the scattering angle in tantalum,
\SI{8}{\percent} below CATIMA, whose radiation length of tantalum
differs from the PDG value used in LADA.  This is agreement between two
implementations of the same model, not an experimental accuracy.

\begin{table}[t]
\caption{\label{tab:verification}Selected verification results: bounds on the
largest deviation over the compared map coefficients up to order $n$,
relative except where a unit is given, for the HHT matrix $R$,
and for charge-state fractions (absolute); RF maps are compared by evaluating them at test points.}
\begin{ruledtabular}
\begin{tabular}{llcl}
Quantity & Reference & $n$ & Agreement\\
\hline
dipole fringe, $y$ block & COSY & 1 & $10^{-8}$\\
bend, fringes, $x,a,y,b$ & COSY & 3 & $5\times10^{-9}$\\
bend, field step & traced rays & 1 & $2\times10^{-8}$\\
multipole fringes, 6D & COSY & 5 & $3\times10^{-6}$\\
el.\ quadrupole fringe, 6D & COSY & 5 & $5\times10^{-5}$\\
the same, pulsed & float Lorentz & 5 & \SI{6e-8}{m}\\
gradient profile, 6D & COSY & 4 & $3\times10^{-4}$\\
bend, Maxwellian edge & float Lorentz & 2 & $3\times10^{-5}$\\
bend edge, gap, $(b|yyy)$ & float Lorentz & 3 & $10^{-2}$\\
ESR, Maxwellian, $\xi_{x,y}$ & PTC & 2 & $10^{-11}$\\
ESR, Maxwellian, $\Delta L/L$ & PTC & 3 & $10^{-5}$\\
dipole, time of flight & COSY & 3 & $3.4\times10^{-8}$\\
quadrupole, time of flight & COSY & 3 & $5.2\times10^{-9}$\\
HHT line, $R$ ($4\times4$)\footnotemark[2] & COSY & 1 & $2\times10^{-13}$\\
HHT line, four $T_{ii6}$\footnotemark[2] & COSY & 2 & $2\times10^{-14}$\\
HHT line, 6D map\footnotemark[3] & COSY & 9 & $5\times10^{-7}$\\
ESR one turn, 6D map\footnotemark[3] & COSY & 5 & $5\times10^{-7}$\\
RF cavity, zero field & exact drift & 3 & $10^{-13}$\\
RF cavity, 6D map & float Lorentz & 3 & $10^{-9}$\\
RF line, rebased & float Lorentz & 3 & $2\times10^{-9}$\\
CH on-axis gain & RF-Track & -- & \SI{7e-5}{keV/u}\\
buncher $R_{21}$ & RF-Track & 1 & $2\times10^{-6}$\\
p in Al, \SI{100}{MeV} & NIST PSTAR & -- & \SI{0.01}{\percent}\\
ions, $\ge\SI{5}{MeV/u}$\footnotemark[1] & CATIMA & -- & \SI{0.6}{\percent}\\
charge fractions & GLOBAL & -- & $4.7\times10^{-6}$\\
charge fractions & CHARGE & -- & $4.8\times10^{-5}$\\
\end{tabular}
\end{ruledtabular}
\footnotetext[1]{Scattering angle in tantalum excepted: \SI{-8}{\percent}, from
CATIMA's radiation length of tantalum.}
\footnotetext[2]{Identical atomic-mass constants in both codes. With their
native constants the bounds are $5\times10^{-6}$ in $R$ and
$4.1\times10^{-7}$ in the four selected $T_{ii6}$, $i=1,\ldots,4$.}
\footnotetext[3]{Every coefficient COSY prints, time row included, at its
printed precision; ESR with COSY's hard edges and sextupoles off.}
\end{table}

The RF-Track rows (Table~\ref{tab:verification}) use the same continuous on-axis field, RF phase and
particle mass in both codes. LADA's smooth reconstruction is sampled
on successively finer grids for RF-Track's linear interpolation;
the quoted residuals use 32-fold grid refinement; a further fourfold
refinement reduces both residuals. CH0 uses a CST-derived
axis profile, while B1 uses its model profile. These tests compare
on-axis acceleration and the canonical coefficient
$R_{21}=\partial(p_x/p_0)/\partial x$, respectively; they do not test
the complete 3D fields. The latter retain azimuthal field components
absent from an axisymmetric reconstruction through $r^2$,
where $r=\sqrt{x^2+y^2}$.
For LADA's hard-edge solenoid, the energy dependence of the transverse
map is checked against the analytic matrix expressed in LADA's
momentum coordinates and against independent integration of the
Lorentz equations through the solenoid field.

\begin{figure*}[!tp]
  \centering
  \includegraphics[width=\textwidth]{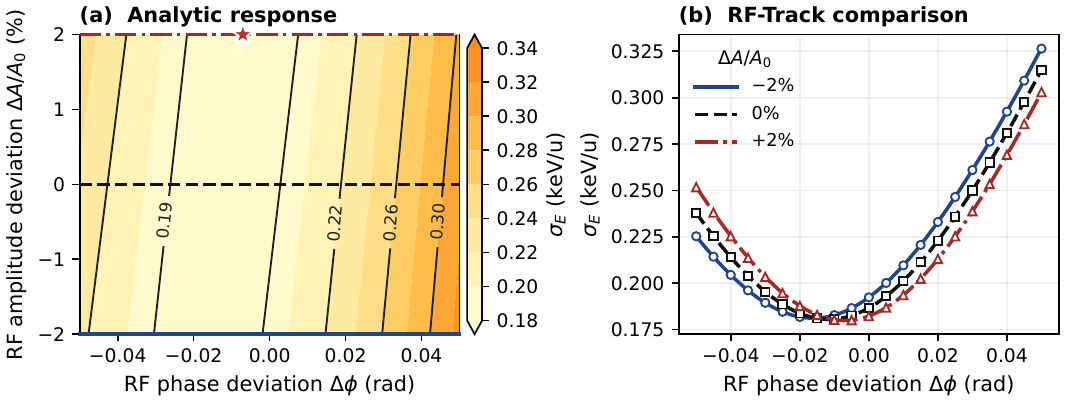}
  \caption{\label{fig:rf_response}Exit-energy spread as a function of
  two RF knobs for the local CH2 test. (a)~Spread evaluated from one
  third-order map and the input moments, Eq.~\eqref{eq:knob_moments};
  horizontal lines: the sections shown in (b); star: minimum within the
  knob ranges. (b)~Sections at three amplitudes: analytic-map curves and
  RF-Track points. The rms difference is below \SI{0.08}{\percent}
  on the full $21\times21$ comparison grid. Input: rms RF phase
  \SI{0.01}{rad}, $\delta=\xi+2\times10^{-5}\phi$, with independent
  Gaussian $\xi$ of rms $2\times10^{-7}$ and 3000 fixed samples;
  transverse coordinates zero. The response polynomial is evaluated
  without transporting particles at each setting.}
\end{figure*}

A local longitudinal test uses the eight-gap CH2 model profile,
$^{48}$Ca$^{8+}$ at \SI{2.3}{MeV/u}, 3000 fixed particles and amplitude
and phase knobs within $\pm2\%$ and $\pm0.05$ rad. The third-order map
contains six phase-space variables and two knobs. Empirical input
moments give the exit-energy variance by Eq.~\eqref{eq:knob_moments}.
Figure~\ref{fig:rf_response} shows the resulting response surface and
sections compared with RF-Track 2.5.5 on a $21\times21$ grid. Both
codes use the same nominal phase, particle sample and continuous
on-axis field; RF-Track uses parallel adaptive Runge--Kutta integration.
The rms spreads agree within \SI{0.08}{\percent} over the grid.
Analytic variance gradients agree with central differences of RF-Track
at three tested points within \SI{1}{\percent}.

The response surface is computed from one map, without a particle
loop at each setting. Table~\ref{tab:rf_fit} separates construction
from reuse. A scan on the prepared polynomial is about $10^4$ times
faster than direct tracking; including map and moment construction
still gives a factor of 83 for this scan.
The minimum of the exit-energy spread within the knob ranges was also
searched for in both codes with the same bounded quasi-Newton optimiser,
L-BFGS-B~\cite{byrd1995lbfgsb}: in LADA on the map polynomial with its analytic gradient, in RF-Track by tracking with
finite-difference gradients (Table~\ref{tab:rf_fit}). The rms spreads at
the two optima agree within $8\times10^{-6}$ relative; the agreement
holds when RF-Track is repeated with a smaller integration step and a
finer field grid. DA integration uses 400 steps; 2400 steps change
the rms by less than $10^{-7}$ at the checked settings. This is a
single-cavity test with zero transverse emittance and no collective
effects, not a timing of the complete SIBAF beam line.

\begin{table}[t]
\caption{\label{tab:rf_fit}Local two-knob CH2 response and optimisation
on the Apple M3 laptop with eight RF-Track threads. Times in seconds;
three fresh map constructions, three scans and nine searches at a fixed
nominal phase, reported as medians. Field and input-beam
preparation are excluded. The independent check tracks the DA optimum
with RF-Track.}
\begin{ruledtabular}
\begin{tabular}{lrr}
Operation & LADA & RF-Track\\
\hline
One variance evaluation & $4.3\times10^{-6}$ & 0.062\\
$21\times21$ response scan & 0.0019 & 27.9\\
Map and moment construction & 0.334 & --\\
Search on prepared model & 0.00036 & 2.35\\
Map, moments, search and check & 0.397 & 2.35\\
\end{tabular}
\end{ruledtabular}
\end{table}

% ---------------------------------------------------------------------------

\section{Migration}
\label{sec:migration}

Existing lattices can be imported from COSY, MAD-X, Xsuite and MIRKO
input.  The import writes the lattice as a plain Python module and a
report that labels every source record as exact, approximated,
diagnostic or unsupported.  The trust pipeline then rebuilds the
lattice, checks symplecticity, runs the source program when it is
installed, and states one verdict that names exactly what was compared.
All 110 records of the HHT final-focus deck in COSY import exactly
(Fig.~\ref{fig:beamlines}b); its
comparison shows that the transverse matrix and four selected
second-order coefficients agree
with the COSY run (Table~\ref{tab:verification}); it does not certify
the higher orders or the longitudinal plane.  The MIRKO export of
the ESR isochronous setting of 2021 imports its quadrupoles, dipoles, pole faces
and repeated elements exactly and reproduces the periodic solution
MIRKO printed into the file: $\gamma_t=1.563$, dispersion
\SI{-4.070}{m} at the cooler, $\beta$ functions within \SI{0.3}{\%}.
Its 62 sextupole records remain approximations: 54 sextupole components
of the dipoles with negligible strength are kept as markers, and the
strength normalisation of the eight discrete sextupoles has not been
checked against MIRKO. A target record has no LADA counterpart, so the
verdict is \texttt{FAIL}.

% ---------------------------------------------------------------------------
\section{Applications}
\label{sec:applications}

\begin{figure*}[!tp]
  \centering
  \includegraphics[width=\textwidth]{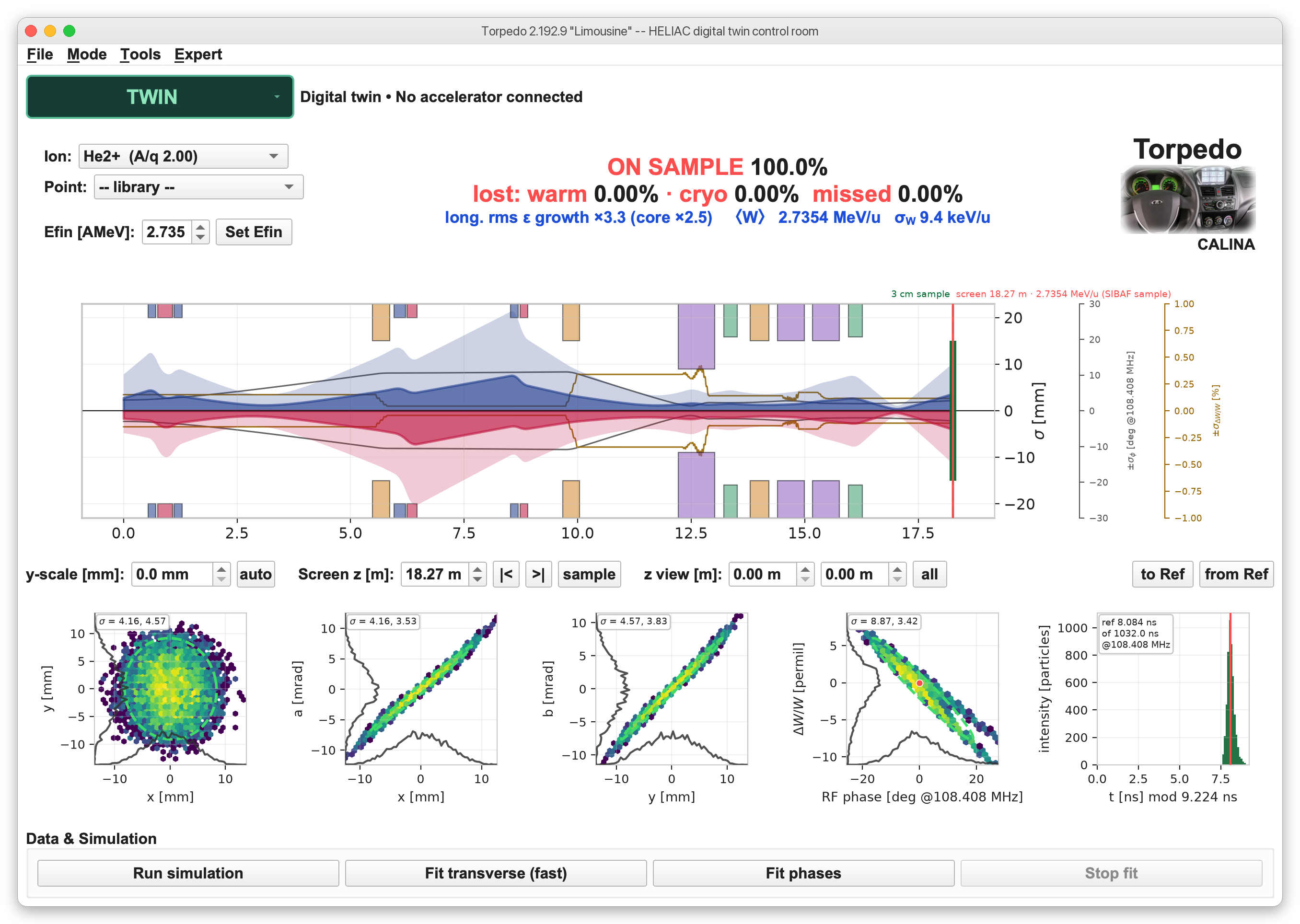}
  \caption{\label{fig:calina}A screenshot of Torpedo, the GUI of the CALINA digital twin
  for SIBAF, rendered with a light print palette. The beam-line view
  shows the injector-to-CM1 transport and the sample plane; the lower
  panels are the transverse spot, two transverse phase portraits,
  longitudinal phase space and arrival-time distribution. The readout
  at the top gives the fraction of ions on the sample, the losses in the
  warm and cryogenic sections and the fraction missing the sample.
  The displayed beam quantities are calculated for the nominal
  He$^{2+}$ startup setting.}
\end{figure*}

\paragraph{SIBAF based on the HELIAC demonstrator.}
The SIBAF project is a new ion-irradiation facility at GSI (a
superconducting heavy-ion linac) for the qualification of fusion-reactor materials. It combines the existing
high-charge-state injector HLI and the first superconducting
cryomodule CM1 of the future HELIAC facility~\cite{schwarz2020heliac,linac2024tuaa004}
with a dedicated sample station.
CALINA uses LADA to model this beam line as a digital twin, using the
layout supplied by the SIBAF team and published cavity
parameters~\cite{schwarz2020heliac}. RF models use CST field exports
where available and model profiles otherwise; the solenoid model uses
Cryomagnetics' calculated field maps, cross-checked against the
manufacturer's axial field measurements during acceptance testing.
Torpedo is CALINA's GUI, intended primarily for use in the SIBAF control
room (Figs.~\ref{fig:linac} and \ref{fig:calina}). Connection to the
hardware is planned through ChASSIS, a custom Python control and
data-acquisition library.

With CALINA v2.192.14 on the same laptop, a quadrupole or solenoid
strength change produces new envelopes and phase portraits in
\mbox{3.98--\SI{4.10}{s}}; a cavity-voltage change takes
\mbox{5.33--\SI{5.37}{s}}. These times include map updates, transport
of 5000 particles at fifth order with 20 steps per RF gap, and display.
A local transverse search over nine knobs takes \SI{14.2}{s} for
three second-order DA sweeps. Each step expresses the beam covariance
matrix as a polynomial in the knobs and minimises the quadratic
constraints on it analytically; particle losses are checked afterwards
by tracking, which is not included in this time.

\begin{table*}[!t]
\caption{\label{tab:calina_timing}CALINA/LADA response times on an
Apple M3 laptop (eight CPU cores, 24 GiB RAM, compiled DA kernel), and
estimated for a high-end workstation. GUI timings run from a changed
setting to completed beam calculation and display. Magnetic and RF updates
are three successive 1\% strength increments from the nominal helium
startup point; unchanged element maps remain cached.}
\begin{ruledtabular}
\begin{tabular}{llll}
Operation & Calculation & Laptop (s) & Workstation estimate (s) \\
\hline
Quadrupole/solenoid update & 5000 particles, order 5 & 3.98--4.10 & 2.0--2.7 \\
RF-voltage update & 5000 particles, order 5 & 5.33--5.37 & 2.7--3.6 \\
RF-knob map construction & 8 variables, order 3 & 0.97 & 0.49--0.64 \\
RF-knob search & 3000 particles, 132 evaluations & 0.67--0.84 & 0.34--0.56 \\
Transverse local search & 9 knobs, 3 order-2 sweeps & 14.2 & 7.1--9.4 \\
Energy change through display & 40000-particle check, order 5 & 16.8 & 7.9--10.5 \\
\end{tabular}
\end{ruledtabular}
\end{table*}

For helium irradiation, a nominal scan from \SI{1.400}{MeV/u} to
\SI{3.255}{MeV/u} keeps each transverse rms spot size at the sample
within \SI{2.6}{\percent} of its value at the top of the scan. These are
calculated targets, not measured operating points yet. Each setting is
checked with LADA's section-wise particle tracking through fifth-order
DA maps of the magnets and of the RF cavities.
Requesting \SI{2.4}{MeV/u} from the stored scan curves takes
\SI{16.8}{s} through to the updated portraits: a 40000-particle
reference and candidate check, followed by a 5000-particle display
calculation, at fifth order and 40 steps per RF gap. All calculations
use LADA; no external tracking code is called.

For a high-end control-room workstation, a conditional scaling estimate
assumes 1.5--2 times faster serial execution and 3--4 times faster
parallel stages, including their overhead. With the measured times
split into serial and parallel parts (Table~\ref{tab:calina_timing}),
this gives approximately 2--\SI{3.5}{s} for an interactive update,
0.35--\SI{0.6}{s} for the fixed-map RF-knob search, and 8--\SI{10.5}{s}
for the complete energy change.

\begin{figure}[!tbp]
  \centering
  \includegraphics[width=\columnwidth]{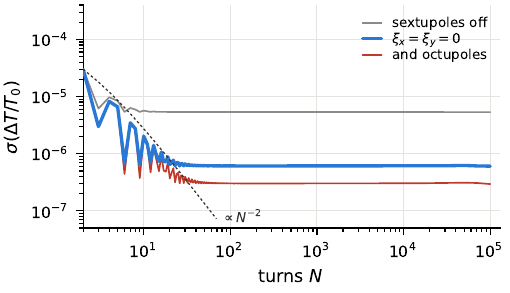}
  \caption{\label{fig:esrturns}ESR, isochronous mode: revolution-time
  spread, the rms width $\sigma$ of $\Delta T/T_0$, of a beam of one species; the
  revolution time of each ion is the slope of a line fitted to its
  arrival times over the first $N$ turns.  1500 ions, rms momentum spread
  $\sigma_{\delta_p}=\SI{0.1}{\percent}$, rms emittance
  $\varepsilon_{\mathrm{rms}}=\SI{5}{mm\,mrad}$ in both planes, order-5
  one-turn map.  Over fewer than ${\sim}50$ turns the betatron oscillation of
  the arrival times has not averaged out and still biases the fitted
  slope: $\propto N^{-2}$ (dashed), modulated by the horizontal tune
  (fractional part 0.35).  Sextupole settings as in
  Fig.~\ref{fig:rings}c (zero betatron amplitude).  With $\xi=0$ the
  width comes from $c_3=-105$ acting on the momentum spread and from
  the second-order chromaticity; octupoles remove the first
  contribution.}
\end{figure}

\paragraph{Proton microscopy.}
The PRIOR-II proton microscope, built for FAIR and commissioned at
HHT~\cite{schanz2024prior}, is modelled at its published working point
(Fig.~\ref{fig:beamlines}a),
with the matching of its second-order chromatic aberrations
$T_{126}$ and $T_{346}$.  PRIORA~\cite{varentsov2026priora}, an averaged
Monte-Carlo code for proton radiography, tracks particles through LADA maps of
arbitrary order and through matter with LADA's stopping and scattering
tables.  A pulsed isotrajectory optics for laser-accelerated proton radiography,
with time-dependent electric quadrupoles, was designed with
LADA as well~\cite{varentsov2026prio}.

\paragraph{Final focus and beam shaping.}
The typical HHT user beam is U$^{73+}$ at \SI{300}{MeV/u}
($B\rho=\SI{8.76}{T\,m}$).  For a given ion and target the specific energy deposited at the
spot centre is proportional to the fluence there per incident ion, $F$.
The settings of all 14 quadrupoles were found in LADA by maximizing $F$ for a
round spot and a limited line loss on linear tracking; every number below is
from a Gaussian beam of the uncooled SIS-18 emittance tracked through all
apertures with third-order maps, where the aberrations of the final triplet
lower $F$ (orders 3, 5 and 7 agree).  The stigmatic $0.9\times\SI{0.9}{mm^2}$
FWHM waist of the linear optics, the smallest round spot, loses half of the
beam in the line; the worst single aperture alone accounts for
\SI{30}{\percent}.  With the loss limited to \SI{25}{\percent} the spot is
$1.02\times\SI{0.96}{mm^2}$ FWHM (Fig.~\ref{fig:beamlines}b) and $F$ is
\SI{12}{\percent} higher.  Fewer losses cost deposition: at
\SI{5}{\percent} the spot grows to $1.5\times\SI{1.4}{mm^2}$ and $F$
falls by \SI{36}{\percent} against the waist.  Without a limit the setting scrapes two thirds of the
beam in the bending magnets for \SI{26}{\percent} more $F$ than the waist.  For
$\varepsilon_n=30\times\SI{8}{mm\,mrad}$ the setting for
\SI{25}{\percent} loss increases $F$ by a factor of 1.53 over the $40\times10$ setting.  A cooled beam of the same intensity
would allow a further optimization of the final focus for
plasma-physics experiments.  U$^{28+}$ at \SI{196}{MeV/u}, more intense
but not coolable, has \SI{18.0}{T\,m}; the same focusing would need
1.45 and 2.1 times the maximum gradients of the last two quadrupoles,
or would result in a much larger focal spot.

High-energy-density experiments at FAIR would benefit from uniform
target irradiation.  With two octupole/duodecapole elements in the FAIR
HEDgeHOB line (Fig.~\ref{fig:beamlines}c--e), LADA computes the moments
of the final distribution from the order-5 map and Gaussian input
moments, and fits the element strengths to the moments of a uniform
distribution, the method of Ref.~\cite{varentsov2005shaping} formerly
run in COSY.  Tracked element by element (300\,000 ions), the fourth normalized moment of the spot
falls from 3.0 to 1.9 in $x$ and 1.95 in $y$ (uniform 1.8); orders 5 and 7 agree.

\paragraph{Mass separators.}
The FRS~\cite{geissel1992frs} is transcribed from the GICOSY deck of
its standard setting (Fig.~\ref{fig:rings}a): length \SI{73.07}{m},
achromatic at S4 ($|R_{16}|<\SI{1}{mm}$).  With hard edges LADA gives
$(x|\delta_p)=\SI{-6.481}{m}$ at S2, against \SI{-6.490}{m} from
GICOSY for the same deck, and $(x|a)=\SI{0.199}{m}$ at S2 and
\SI{0.141}{m} at S4, against \SI{0.132}{m} and \SI{-0.016}{m}; the
difference is the fringe fields of the GICOSY model, which are
point-symmetric about the effective boundary.  Integrated through
fringe fields of logistic shape, $1/(1+e^{z/w})$, whose widths $w$
reproduce the GICOSY fringe-field integrals, the first-order coefficients at S1--S4 agree to \SI{2.1}{mm}
in $(x|a)$, \SI{0.09}{mm} in $(y|b)$ and $2\times10^{-4}$ in $(x|x)$, and
$(x|\delta_p)$ at S2 is \SI{-6.4905}{m} against \SI{-6.4903}{m}.  The
remaining $(x|a)$ follows from a difference of $9\times10^{-6}$ in
$(a|x)$ of each dipole between this integration and the fringe-field
integral method of GICOSY.  Higher orders were not compared.  A model
lattice of the Super-FRS (second-order separator correction, fragment
selection with fragmentation and charge states) is a part of the
LADA distribution's examples.

\paragraph{Synchrotrons and storage rings.}
In the isochronous mode of the
ESR~\cite{hausmann1998esr} the revolution time of the stored ions does
not depend on their momentum to first order; the isochronous condition
is $\gamma=\gamma_t$.  The imported setting of 2021 has
$\gamma_t=1.563$ and a dispersion of \SI{-4.07}{m} in both straight
sections (Fig.~\ref{fig:rings}b); LADA sets the reference ion to the
$\gamma_t$ of the one-turn map.  On the closed orbit
$\Delta T/T_0=-\delta f/f=c_2\delta_p^2+c_3\delta_p^3+\dots$; second- and
third-order isochronicity mean $c_2=0$ and $c_3=0$.

\emph{Edge fields.}  Dipoles and quadrupoles have Maxwellian edges
(Sec.~\ref{sec:code}), the dipoles with the fringe integral of the deck.
They decide the result.  COSY's hard edge gives $\xi_y=-2.75$ for the
sextupoles off, the Maxwellian edge $-4.34$; $\xi_x=-4.52$ is the same.
The quadrupole edges move $c_3$ by $-97$, from $+34$ to $-62$ with the
sextupoles off.  PTC gives the same $\xi_{x,y}$ and, with its quadrupole
fringe switched on, the same path length to third order
(Table~\ref{tab:verification}; compared without the fringe integral of the
dipoles).  With Enge fringes of the quadrupoles instead, over full
apertures of 0.02--\SI{0.36}{m}, $c_3$ lies between $-59$ and $-69$.

\emph{Second order.}  $c_2$ is linear in the strengths $k_2L$ of the eight
sextupoles.  Equal strengths within each mirror pair give
$\xi_x=\xi_y=0$ and $c_2=-0.021$, the correction of Ref.~\cite{hausmann1998esr}; $\xi_x$, $\xi_y$ and $c_2$ are not
independent in this mirror-symmetric optics.

\emph{Third order.}  At this setting $c_3=-105$, against $-8$ without the
quadrupole edges.  $c_3$ is a quadratic function of the $k_2L$ and can be searched
numerically.  The sextupoles alone reach $c_2=c_3=0$, with $\xi\ne0$; the weakest setting found by a numerical
search has 21 times the strength of the second-order correction, and $\xi_x=\xi_y=c_3=0$ needs 22 times, with strong pairs of opposite sign.  Several sign
arrangements have the same strength; the beam of Fig.~\ref{fig:esrturns} loses 8--\SI{87}{\percent} of its ions
within 300 turns (exact sextupole kicks, section maps of order 5 and 7), depending on the arrangement, and the rest has
$\sigma\approx1$--$2\times10^{-5}$.  They are not a working setting.  Thin
octupoles beside the sextupoles (hypothetical) would give
$c_3=0$ without changing $\xi$ and $c_2$.
$|\delta f/f|$ at $\delta_p=\pm0.2\,\%$ falls from $1.6\times10^{-5}$
(sextupoles off) to $8.9\times10^{-7}$, and to $8\times10^{-8}$ with the
octupoles (Fig.~\ref{fig:rings}c).

\emph{Revolution-time spread.}  The beam of Fig.~\ref{fig:esrturns} is tracked over
$10^5$ turns with the order-5 one-turn map.  It loses 0.4--\SI{0.5}{\percent} of the
ions with the sextupoles on, none before $3\times10^4$ turns; tracked with exact sextupole
kicks and section maps of order 5 and 7, the same ions stay, so this is a truncation effect.  With $\xi=0$ the revolution-time
spread settles at $\sigma\approx6\times10^{-7}$ within 100 turns.  The momentum
spread alone gives $3.5\times10^{-7}$, the cubic term cut at
$3\sigma_{\delta_p}$; the emittance alone $4\times10^{-9}$.  The rest is
the product of both, the momentum dependence of the chromaticity; it also
comes from the quadrupole edges, without which $\sigma=3.7\times10^{-8}$.
The octupoles reduce $\sigma$ to $3.0\times10^{-7}$.  An offset of
$\gamma$ from $\gamma_t$ adds $c_1\delta_p$, which partly cancels
$c_3\delta_p^3$: for a beam of 6000 ions over 1000 turns, with all fields
scaled to the ion's rigidity, the spread is
smallest at $\Delta\gamma/\gamma=+7\times10^{-4}$, less than half of that at
$\gamma_t$.  The mass resolving power
$m/\Delta m=1/(2.355\,\gamma_t^2\sigma)$ is about $3\times10^5$ at $\gamma_t$
and $6\times10^5$ at the offset or with the octupoles.  It is a property of
this beam and of the edge model: with Enge fringes of 0.1, 0.2 and
\SI{0.3}{m} aperture $\sigma$ is $6.0$, $6.6$ and $8.9\times10^{-7}$; the real aperture of the
quadrupoles was not available.  Versions 1 and 2 of this article left these
edge terms out; version 2 gave $c_3=-7.6$, $\sigma=2.6\times10^{-8}$ and $m/\Delta m\approx6\times10^6$.
Model lattices of the ESR and the FAIR Collector Ring (isochronous
settings, multi-turn tracking) are part of the LADA distribution's examples.

\paragraph{Light and lasers.}
For the PHELIX petawatt laser~\cite{major2024phelix} LADA computes the
final focus, corrects a measured wavefront and shapes the focal spot by
the moment method used for HEDgeHOB: phase terms
$A_4(r/R)^4+A_6(r/R)^6$ before an f/1.33 parabola are fitted to the
second and fourth moments of the spot.  For a flat-top beam the
geometric rms spot radius falls from 5.2 to \SI{0.22}{\micro m}; the
terms fitted for a Gaussian beam leave \SI{0.42}{\micro m}.  A chirped
pulse maps frequency to arrival time, so a static spectral phase mask
corrects transverse chromatic aberrations synchronously: for the PHELIX
spectrum (\SI{1053}{nm}, \SI{9}{nm} FWHM) at f/1.5 it restores the
diffraction-limited spot behind an assumed focusing optic whose lateral
colour and chromatic defocus widen it to 1.9 times the diffraction limit.

% ---------------------------------------------------------------------------
\section{Summary and limits}
\label{sec:summary}

LADA combines DA transfer maps, extended RF fields, ion--matter
interaction models and loss-aware particle tracking in Python. Knob maps yield
explicit response functions and analytic sensitivities; input-moment
contraction removes repeated particle transport from smooth beam-objective
scans and matching. Setup, reuse and independent verification have
distinct costs (Table~\ref{tab:rf_fit}).
Space charge, plasma and wake fields are excluded. Transfer maps are
local expansions; their validity is checked by comparing orders and
by direct integration.

\section*{Code availability and development}
The code and calculation scripts are available from the authors.
Project-authored code is licensed under MIT. An AI agent assisted
with the coding under D.V.'s guidance and control; physics changes require the verification
of Sec.~\ref{sec:verification}. The calculations and numerical results
presented here correspond to LADA~v1.11.13.

% ---------------------------------------------------------------------------

\end{document}